\documentclass[11pt,twoside]{article}
\usepackage{amsmath}
\usepackage{amssymb}
\usepackage{cite}
\usepackage{amsfonts}
\usepackage{graphicx}

\begin{document}
\newcommand{\pst}{\hspace*{1.5em}}

\newcommand{\rigmark}{\em Journal of Russian Laser Research}
\newcommand{\lemark}{\em Volume 30, Number 5, 2009}

\newcommand{\be}{\begin{equation}}
\newcommand{\ee}{\end{equation}}
\newcommand{\bm}{\boldmath}
\newcommand{\ds}{\displaystyle}
\newcommand{\ba}{\begin{eqnarray}}
\newcommand{\ea}{\end{eqnarray}}
\newcommand{\baa}{\begin{array}}
\newcommand{\eaa}{\end{array}}
\newcommand{\arcsinh}{\mathop{\rm arcsinh}\nolimits}
\newcommand{\arctanh}{\mathop{\rm arctanh}\nolimits}
\newcommand{\bc}{\begin{center}}
\newcommand{\ec}{\end{center}}

\thispagestyle{plain}

\label{sh}


\begin{center} {\Large \bf
\begin{tabular}{c}
DESCRIPTION
\\[-1mm]
OF CLASSICAL AND QUANTUM INTERFERENCE
\\[-1mm]
IN VIEW OF THE CONCEPT OF FLOW LINE
\end{tabular}
 } \end{center}

\bigskip

\bigskip

\begin{center} {\bf
Milena Davidovi\'c,$^{1}$ \'Angel S. Sanz,$^{2}$ and Mirjana Bo\v zi\'c$^{3*}$
}\end{center}

\medskip

\begin{center}
{\it
$^1$Faculty of Civil Engineering, University of Belgrade,
11000 Belgrade, Serbia

\smallskip

$^2$Instituto de F{\'\i}sica Fundamental (IFF-CSIC),
Serrano 123, 28006 Madrid, Spain

\smallskip

$^3$Institute of Physics, University of Belgrade,
11080 Belgrade, Serbia
}
\smallskip

$^*$Corresponding author e-mail:~bozic@ipb.ac.rs\\
E-mails: milena@grf.bg.ac.rs, asanz@iff.csic.es
\end{center}

\begin{abstract}\noindent
Bohmian mechanics, a hydrodynamic formulation of quantum mechanics,
relies on the concept of trajectory, which evolves in time in
compliance with dynamical information conveyed by the wave function.
Here this appealing idea is considered to analyze both classical and
quantum interference, thus providing an alternative and more
intuitive framework to understand the time-evolution of waves,
either in terms of the flow of energy (for mechanical waves, sound
waves, electromagnetic waves, for instance) or, analogously, the
flow of probability (quantum waves), respectively. Furthermore, this
procedure also supplies a more robust explanation of interference
phenomena, which currently is only based on the superposition
principle. That is, while this principle only describes how
different waves combine and what effects these combinations may lead
to, flow lines provide a more precise explanation on how the energy
or probability propagate in space before, during and after the
combination of such waves, without dealing with them separately
(i.e., the combination or superposition is taken as a whole). In
this sense, concepts such as constructive and destructive
interference, typically associated with the superposition principle,
physically correspond to more or less dense swarms of (energy or
probability) flow lines, respectively. A direct consequence of this
description is that, when considering the distribution of
electromagnetic energy flow lines behind two slits, each one covered
by a differently oriented polarizer, it is naturally found that
external observers' information on the slit crossed by single
photons (understood as energy parcels) is totally irrelevant for the
existence of interference fringes, in striking contrast with what is
commonly stated and taught.
\end{abstract}

\medskip

\noindent{\bf Keywords:}
Bohmian mechanics, flow line, interference, sound wave, electromagnetic
wave, Umov vector, Poynting vector, probability current


\section{Introduction}
\label{sec1}
\pst
Superposition and interference are two intertwined capital concepts
in both classical and quantum wave mechanics. As is well known,
within the framework of a general wave theory, superposition simply
consists in adding two or more waves. In elementary courses on wave
mechanics \cite{ref1,ref2,ref3}, it is commonly taught that when two
traveling waves propagate through a certain medium, the net
displacement of the medium at a particular point and a given time is
just the direct sum of the individual displacements associated with
each contributing wave.

Such a traditional conception of interference, where only the net
effect of the waves is considered and the associated flow of energy
is often neglected (or just pushed into the background), strongly
relies on the mathematical grounds of wave mechanics, namely the
superposition principle. According to this principle, given a set of
field-like variables $\{ \vartheta_i \}$, solutions to a linear flow
equation
\be
 \Phi \vartheta_i = 0 ,
 \label{eq1}
\ee
where $\Phi$ denotes the (space and time) flow operator
\be
 \Phi = \frac{\partial^2}{\partial t^2}
   - c^2 \frac{\partial^2}{\partial x^2} ,
 \label{eq2}
\ee
and $c$ is a constant typically identified with the propagation (or
diffusion) speed of the field variable, new solutions can be readily
built up from linear combinations of the $\vartheta_i$,
\be
 \Theta = \sum_{i=1}^N c_i \vartheta_i .
 \label{eq3}
\ee
This intuitive principle has an important practical advantage: it
allows us to find and devise smart methods to solve linear flow
equations, like Eq.~(\ref{eq1}), by just decomposing the total
solution into a set of partial contributions. Besides, in our
opinion it also brings in an important inconvenience: our conception
of physical superposition is directly grounded on its mathematical
basis, which becomes critical when Eq.~(\ref{eq1}) represents a wave
equation and the field variables refer to waves propagating from
different sources. As a consequence, for example, following the
above mathematical roots, it is commonly thought that when two waves
superimpose, they pass each other just as if each wave was unaware
of the other; only the net effect matters \cite{ref2}. This is just
a physical model to understand and explain the wave phenomena
observed in Nature. However, it also strongly determines our
perception of such phenomena, what has very important consequences
particularly in the case of quantum systems.

To some extent, such a picture of interference is rather poor. It
refers to the superposition or combination of partial waves,
although in real life these waves never appear separately. In order
to get some more light on interference, let us get back to the usual
notion of wave from classical wave theory, where it typically
represents a perturbation that propagates throughout a certain
medium. From a physical viewpoint, the propagation of a perturbation
in a material medium is associated with a flow of energy from one
place to another within such a medium. A better understanding of
interference should therefore include the tools to determine how
this flow takes place, and not only rely on how different wave
components (which we never observe in real life) combine. The same
idea, of course, can be naturally extended to quantum systems. Even
if the nature of quantum waves is different from that of classical
waves, the key element, namely the transport or flow of a certain
quantity (probability in the case of the former and energy for the
latter) is the same.

In this work we tackle the issue of interference within a general
theoretical framework based on the concept of flow line, applicable
to both classical and quantum systems. Within this scenario, while
the superposition principle only tells us how different waves
combine, the trajectories or flow lines provide us with information
on how the net effect propagates before, during and after their
combination, putting the emphasis on the more natural conception of
considering the wave as a whole, and not on splitting it up in its
different components, as it is typically done.

It is worth noting that, although the idea of transversal flow is
mentioned in different sources (for example, see energy flow for
non-dispersive waves in \cite{ref1}, or for the probability flow in
quantum processes in \cite{ref4,ref5}), the idea of monitoring it
during the full propagation of the wave (i.e., time by time) is not
that general at all. As far as we know, the use of flow lines as a
working tool to visualize the propagation of energy in sound waves
can be traced back to the 1980s \cite{ref6,ref7,ref8,ref9,ref10}.
The concept of flow line is quite general and can be found in many
different physical contexts as well \cite{ref11}, including
classical electromagnetic problems (see Ref.~\cite{ref12} and
references therein). Specifically, in most of these works this
concept is introduced after assuming an analogy between the
corresponding problem and a hydrodynamic one.

Together with the fact that flow lines constitute an interesting
working tool to analyze the evolution (flow) of energy or
probabilities, we would also like to stress that depending on the
problem considered, these elements present the nice feature that
they can be somehow inferred from experimental data, as shown
recently in the case of Young's two-slit experiment \cite{ref13}.
This representation is in compliance with the recent method devised
to determine the photon wave function \cite{ref14}, based on the
so-called weak-measurement technique \cite{ref15}, which constitutes
a remarkable alternative to the more traditional construction of
phase-space tomograms \cite{ref16}.

This paper is organized as follows. In Sec.~\ref{sec2} we
discuss sound waves as an example of classical waves, introducing
the associated equation for the energy flow lines. As an
illustration, we present a simulation of the energy flow lines
behind a wall with two openings. In this example, the nodes in the
eventual fringe pattern represent regions of silence (or, to be more
precise, low sound) in between regions where the sound is
reinforced. This experience could be equally reproduced with two
synchronized loudspeakers emitting the same sound pitch. The
homologous equation for electromagnetic energy (EME) flow lines is
given in Sec.~\ref{sec3}. Moreover, in analogy with the previous
case, in this section we also show the EME flow lines behind two
slits, although each slit is followed by a linear polarizer with its
polarization axis oriented in a different direction. In Sec.~\ref{sec4}
we discuss how an even more general picture of
interference can be achieved for quantum nonzero-mass particles
after considering flow lines of quantum-mechanical probability
density. Finally, the main conclusions arising from this work are
summarized in Sec.~\ref{sec5}.


\section{Sound Energy Flow Lines: Interference behind Two Openings}
\label{sec2}

In order to derive the equation of motion accounting for the
transport of energy of a sound wave propagating through a
non-viscous fluid, we start from the continuity and momentum
equations \cite{ref3,ref17,ref18},
\ba
 \frac{\partial \rho}{\partial t} +
  \nabla \left( \rho {\bf v} \right) = 0 ,
 \label{eq4} \\
 \rho \ \! \frac{\partial {\bf v}}{\partial t} +
  \nabla p + \rho {\bf v} \cdot \nabla {\bf v} = 0 .
 \label{eq5}
\ea
respectively, where $\rho({\bf r},t)$ is the fluid density, ${\bf
v}({\bf r},t)$ is the velocity of a fluid element, and $p({\bf
r},t)$ is the pressure. Assuming the velocity of the fluid and the
fluctuations of the density and pressure are small ($\rho = \rho_0 +
\rho'$, $\rho' \ll \rho_0$, $p = p_0 + p'$, $p' \ll p_0$, ${\bf v} =
{\bf v}'$), after linearization the above equations become
\ba
 \frac{\partial \rho'}{\partial t} + \rho \nabla {\bf v}' = 0 ,
 \label{eq6} \\
 \rho_0 \ \! \frac{\partial {\bf v}'}{\partial t} + \nabla p' = 0 .
 \label{eq7}
\ea
Taking into account the relationship between pressure and density,
\be
 \left.
 \frac{\partial p}{\partial \rho} \right\arrowvert_{\rho = \rho_0}
  = \frac{p'}{\rho'} = c^2 .
 \label{eq8}
\ee
where $c$ is the speed of sound in the corresponding medium, from
Eqs.~(\ref{eq6}) and (\ref{eq7}) we readily obtain the wave equation
for the acoustic pressure,
\be
 \frac{1}{c^2}\frac{\partial^2 p'}{\partial t^2}
  - \nabla^2 p' = 0 .
 \label{eq9}
\ee
The energy transport in the sound field is described by the equation
\be
 \frac{\partial}{\partial t}
  \left( \frac{1}{2} \ \! \rho_0 v'^2 + \frac{p'^2}{2\rho_0 c^2}
   \right) + \nabla \left( p' {\bf v}' \right)  = 0 ,
 \label{eq10}
\ee
where
\be
 {\bf S} = p' {\bf v}'
 \label{eq11}
\ee
is the so-called Umov energy flow vector \cite{ref3,ref19}, the
mechanical equivalent of the Poynting or Poynting-Heaviside vector,
commonly used in electromagnetism \cite{ref20}.

Let us now consider without loss of generality and for simplicity a
sound wave of constant angular frequency $\omega$ (the analysis
could be carried out as well with non-monochromatic waves). If the
pressure and velocity are taken as complex quantities (a typical
working technique when dealing with wave equations, although only
the real part of these quantities is physically meaningful),
\be
 p' = P e^{-i\omega t} , \qquad {\bf v}' = {\bf V} e^{-i\omega t} ,
 \label{eq12}
\ee
after substitution into Eqs.~(\ref{eq7}) and (\ref{eq8}), we have
\ba
 \nabla^2 P + \left( \frac{\omega}{c} \right)^2 P = 0 ,
 \label{eq13} \\
 i\omega {\bf V} = \frac{\nabla P}{\rho_0} .
 \label{eq14}
\ea
In the case of high-frequency waves, to avoid fast oscillations, we
can consider time-averaging over a cycle, which allows us to
introduce the definition of the time-averaged energy flow vector,
\be
 \langle {\bf S} \rangle = \frac{1}{2} \ \!
  {\rm Re} \left\{ P {\bf V}^* \right\} .
 \label{eq15}
\ee
The modulus of this vector determines the sound intensity
\cite{ref17},
\be
 I = \left\arrowvert \langle {\bf S} \rangle \right\arrowvert ,
 \label{eq16}
\ee
related to the energy density $w$ as
\be
 w = \frac{I}{c} = \frac{|\langle {\bf S} \rangle|}{c} .
 \label{eq17}
\ee

Consider now the propagation of the above monochromatic plane sound
wave incident onto an obstacle with two slits, located on the $XZ$
plane. Assuming the amplitude of the incident complex pressure does
not depend on the z coordinate and has the functional form
$P(x)e^{iky}$, the solution of the Helmholtz equation (\ref{eq13})
behind the obstacle can be expressed in terms of the well-known
Fresnel-Kirchhoff integral from the theory of propagation of optical
waves \cite{ref20,ref21,ref22},
\be
 P(x,y) = e^{-i\pi/4} e^{iky} \sqrt{\frac{k}{2\pi y}}
  \int_{-\infty}^\infty P(x',0) e^{ik(x-x')^2/2y} dx' .
 \label{eq18}
\ee
Here, $P(x',0)$ is the initial pressure, which in our case is given
by the pressure just behind the obstacle, determined by the
corresponding boundary conditions. From Eqs.~(\ref{eq14}) and
(\ref{eq15}), we thus obtain
\be
 \langle {\bf S} \rangle = - \frac{1}{2\rho_0\omega}\
 {\rm Im} \left\{ P\ \! \frac{\partial P^*}{\partial x}\ \! {\bf e}_x
 + P\ \! \frac{\partial P^*}{\partial y}\ \! {\bf e}_y \right\} ,
 \label{eq19}
\ee
where ${\bf e}_x$ and ${\bf e}_y$ are unit vectors along the $x$ and
$y$ directions. The vector field $\langle {\bf S} \rangle$ is
tangent to the sound energy flow lines at each point determined by
the equation
\be
 \frac{d{\bf r}}{dr} = \frac{\langle {\bf S} \rangle}{I} .
 \label{eq20}
\ee
Behind a one-dimensional grating, this equation can be recast as
\be
 \frac{dy}{dx} = \frac{\langle S_y \rangle}{\langle S_x \rangle} .
 \label{eq21}
\ee

\begin{figure}[t]
 \begin{center}
 \includegraphics[width=8cm]{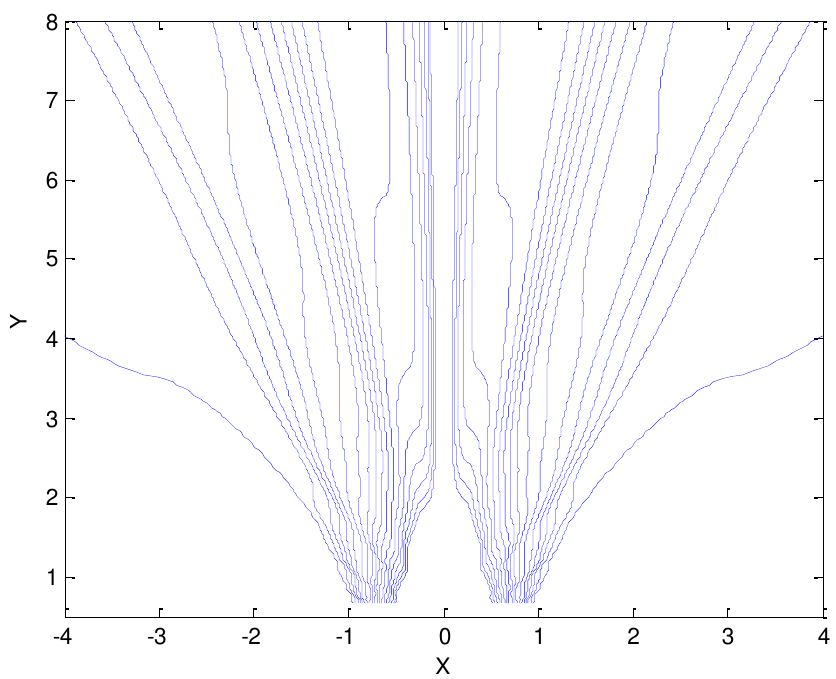}
 \caption{Energy flow lines associated with sound waves emerging from
  an obstacle with two slits.
  The width of the openings is $\delta = 0.5$~m, their mutual distance
  $d = 1.5$~m, the speed of sound $c = 340$~m/s, and its frequency
  $f = 1$~kHz.}
 \label{fig1}
 \end{center}
\end{figure}

Sound energy flow lines calculated from
Eqs.~(\ref{eq18})-(\ref{eq21}) for a totally absorptive obstacle
with two identical openings are shown in Fig.~\ref{fig1}. We find
that the energy flow lines associated with different slits do not
cross the system symmetry line. This means that there is a kind of
fictitious barrier between the two sound sources (slits). The
flux-line non-crossing property for classical/sound waves is
analogous to the Bohmian non-crossing property \cite{ref23,ref24a,ref24b}.
This correspondence can be set on the basis of the analogy between
Eq.~(\ref{eq20}) and the equation of trajectories determined by the
quantum probability current density (see Sec.~\ref{sec4} for a more
detailed discussion).

It is worth noticing that sound waves are described by field
variables, i.e., continuous functions of space and time, although
the gases, liquids or solids through which these waves propagate
consists of individual, discrete objects (e.g., atoms, ions,
molecules, etc.). However, in the case of fluids, for instance,
instead of appealing to these objects to analyze the corresponding
hydrodynamics, it is common to make use of tracer particles, which
help to elucidate the flow dynamics and hence the dynamical
properties of the fluid \cite{ref25,ref26}.

The propagation of sound waves (energy) is of practical interest,
particularly in room acoustics \cite{ref27}. In this area, sound
energy flow lines have been evaluated in the past using different
approximate methods. Recently, the diffraction of these waves at
half planes and slits has been studied \cite{ref28,ref29} by means
of methods of scalar optics in the far field. The method described
above also makes use of results from optics (in general,
electromagnetism), although is more general, since it encompasses
both near and far field.


\section{Electromagnetic Energy Flow Lines: Interference behind Two
Slits Covered by Polarizers}
\label{sec3}

Can the electromagnetic field associated with light be thought as a
kind of ``wave function for the photon''? Although this is a
longstanding challenging question, it is still of much interest
nowadays. Scully and Zubairy summarized \cite{ref30} the objections
posed by Kramers \cite{ref31}, Power \cite{ref32} and Bohm
\cite{ref33} against the concept of a photon, agreeing with Kramers
and Bohm that the concept of a photon wave function should be
carefully used, since it can be very misleading. Nonetheless, they
concluded that each objection can be overcome, suggesting a path to
circumvent them, and providing a way to construct the photon wave
function through the radiation-field second quantization
formulation.

Raymer and Smith \cite{ref34}, Bialinicki-Birula \cite{ref35a,ref35b,ref35c} and
Holland \cite{ref36} went a step further. Holland \cite{ref36}
analyzed similarity of Maxwells' equation with quantum wave
equations (Schrodinger's and Dirac's). Raymer and Smith \cite{ref34}
and Bialinicki-Birula \cite{ref35a,ref35b,ref35c} consider that the usual
electromagnetic Maxwell field is the quantum wave function of the
single photon. The fact that it transforms like a three-dimensional
vector arises from the spin-one nature of the photon. The
interpretation of the Maxwell field is, therefore, akin to the
Schrödinger wave function, which describes the evolution of
probability amplitudes associated with various possible quantum
events in which a particle's position is found to be within a
certain volume. In this sense, Maxwell's equations rule the
evolution of probability amplitudes for various possible quantum
events in which the photon's energy is found within a certain
volume.

Assuming this interpretation for the Maxwell field, and based on a
previous work by Prosser \cite{ref37a,ref37b}, a method to determine the
electromagnetic energy flow lines behind various gratings
illuminated by a monochromaticbeam of light was developed in
\cite{ref12,ref38}. In this case, the solutions of Maxwell's
equations behind the gratings were determined using the solution of
the Helmholtz equation. (It is well known that the space-dependent
part of each component of the electric ${\bf E}({\bf r})$ and
magnetic ${\bf H}({\bf r})$ free fields satisfy Helmholtz's
equation).

As before, again here we are going to consider the time-averaged
energy flux vector to determine the electromagnetic energy flow
lines (for a non-averaged, time-dependent application, see the work
by Chou and Wyatt \cite{ref39}), given by the real part of the
complex Poynting vector \cite{ref40},
\be
 {\bf S}({\bf r}) = {\rm Re} \left[
  \frac{1}{2}\ \! {\bf E}({\bf r}) \times {\bf H}^*({\bf r}) \right] .
 \label{eq22}
\ee
Since the energy flow is along the direction of the Poynting vector,
the EME flow lines will be determined from the parametric
differential equation
\be
 \frac{d{\bf r}}{ds} = \frac{{\bf S}({\bf r})}{cU({\bf r})} ,
 \label{eq23}
\ee
where $s$ is a certain arc-length and $U({\bf r})$ is the
time-averaged electromagnetic energy density
\be
 U({\bf r}) = \frac{1}{4} \left[
    \epsilon_0 {\bf E}({\bf r}) \cdot {\bf E}^*({\bf r})
  + \mu_0 {\bf H}({\bf r}) \cdot {\bf H}^*({\bf r}) \right] .
 \label{eq24}
\ee
As mentioned in the previous section, the Poynting vector in
electromagnetism plays the same role as the Umov vector (\ref{eq19})
in the theory of mechanical waves. Since in the last instance both
vectors are just transport vectors, some authors use the combined
denomination Poynting-Umov vector \cite{ref3} (or also
Heaviside-Poynting-Umov vector, to include Heaviside's
contribution).

Holland argued \cite{ref36} that, for reasons of compatibility with
quantum mechanics, it does not seem to be reasonable assuming that
the tracks ${\bf r} = {\bf r}(t)$ deduced from the equation
essentially equivalent to Eq.~(\ref{eq23}) can be identified with
the orbits of "photons". However, Ghose et al. showed \cite{ref41}
that Bohmian trajectories for relativistic bosons, and so for
photons, could be defined indeed. They derived an equation of motion
for massless bosons, which is equivalent to Eq.~(\ref{eq23}). From
such an equation they obtained photon trajectories behind the
two-slit grating, assuming that the wave function at the point
$(y,x)$, at the sufficient distance $D \gg d^2/\lambda$ to the right
of the plane of the slits, is a superposition of two spherical
waves.

EME flow lines evaluated from Eq.~(\ref{eq23}) in \cite{ref12,ref38}
show that the energy redistribution behind a multiple slit grating
corresponds with a Talbot pattern in the near field, and with a
Fraunhofer interference pattern in the far field. The Fresnel-Arago
laws as well as the Poisson-Arago spot were also interpreted using
EME flow lines \cite{ref42,ref43}. The main conclusion from the
analysis of these phenomena is that the motion of an eventual photon
wave packet essentially represents the flow of electromagnetic
energy along an ensemble of flow lines.

Actually, EME flow lines obtained \cite{ref43,ref44} from a
numerical simulation of Young's two-slit experiment with parameters
taken from the experiment performed by Kocsis et al. \cite{ref13}
showed a good agreement with the averaged photon trajectories
inferred from the experimental data. It is worth stressing that the
photon trajectories reconstructed from the experiment were in
compliance with the Bohmian approach, thus confirming that
trajectories coming from different slits do not cross. This means
that, at the level of the average electromagnetic field (or the wave
function, in the case of material particles, in general), full
which-way information can still be inferred without destroying the
interference pattern. That is, rather than complementarity, the
experiment seems to suggest that a photon wave function has a
tangible (measurable) physical reality \cite{ref44}, in agreement
with a recent theorem on the realistic nature of the wave function
\cite{ref45}.

Let us now evaluate the EME flow lines behind two slits in a more
general case, when each slit is covered by a polarizer with its
polarization axis oriented at a different angle. In our model, we
consider a monochromatic electromagnetic wave in vacuum, traveling
along the $y$ direction, incident onto a two-slit grating located on
the $XZ$ plane, at $y = 0$, with infinitely long slits parallel to
the $z$ axis, so that both the electric and magnetic fields do not
depend on the $z$ coordinate. The electric and magnetic fields of
the incident wave are given by the expressions
\ba
 {\bf E}_{\rm inc} & = & A e^{iky} {\bf e}_z
  - B e^{i\varphi} e^{iky} {\bf e}_x ,
 \label{eq25} \\
 {\bf H}_{\rm inc} & = & \sqrt{\frac{\epsilon_0}{\mu_0}}\ \!
  B e^{i\varphi} e^{iky} {\bf e}_z
  + \sqrt{\frac{\epsilon_0}{\mu_0}}\ \! A e^{iky} {\bf e}_x .
 \label{eq26}
\ea
Since the slits are followed by polarizers, with their polarization
axis at angles $\theta_1$ and $\theta_2$ with respect to the $z$
axis, the $z$ components of the fields behind the grating read as
\ba
 {\bf E}_z({\bf r}) & = & A\cos^2 \theta_1 \psi_1({\bf r})
  + A\cos^2 \theta_2 \psi_2({\bf r}) \nonumber \\ & &
  - B e^{i\varphi} \sin \theta_1 \cos \theta_1 \psi_1({\bf r})
  - B e^{i\varphi} \sin \theta_2 \cos \theta_2 \psi_2({\bf r}) ,
 \label{eq27} \\
 {\bf H}_z({\bf r}) & = & - \sqrt{\frac{\epsilon_0}{\mu_0}}\ \!
    A\cos \theta_1 \sin \theta_1 \psi_1({\bf r})
  - \sqrt{\frac{\epsilon_0}{\mu_0}}\ \!
    A\cos \theta_2 \sin \theta_2 \psi_2({\bf r}) \nonumber \\ & &
  + \sqrt{\frac{\epsilon_0}{\mu_0}}\ \!
    B e^{i\varphi} \sin^2 \theta_1 \psi_1({\bf r})
  + \sqrt{\frac{\epsilon_0}{\mu_0}}\ \!
    B e^{i\varphi} \sin^2 \theta_2 \psi_2({\bf r}) .
 \label{eq28}
\ea
Where $\psi_1({\bf r})$ and $\psi_2({\bf r})$ are scalar functions
that satisfy the Helmholtz equation and the boundary conditions at
the slits. The functions $\psi_1({\bf r})$ and $\psi_2({\bf r})$ can
be recast in the form of a Fresnel-Kirchhoff integral,
\be
 \psi_i(x,y) = \frac{k}{2\pi y}\ \! e^{-i\pi/4} e^{iky}
  \int_{-\infty}^\infty \psi_i(x',0^+) e^{ik(x-x')^2/2y} dx' ,
 \label{eq29}
\ee
%
with $i=1,2$, and where the $\psi_i(x',0^+)$ denote their value just
behind the first and the second slits, respectively, and in front of
the polarizers.

Since the electric and magnetic fields do not depend on the $z$
coordinate,  from Maxwell's equations one obtains two independent
sets of equations \cite{ref20}: one involving the $H_x$ and $H_y$
components of the magnetic field and the $E_z$ component of the
electric field (commonly referred to as $E$-polarization),
\be
 \frac{\partial E_z}{\partial y} = \frac{i\omega}{\epsilon_0 c^2}
  \ \! H_x , \qquad
 \frac{\partial E_z}{\partial x} = - \frac{i\omega}{\epsilon_0 c^2}
  \ \! H_y ,
\label{eq30}
\ee
and another involving $E_x$, $E_y$ and $E_z$ ($H$-polarization),
\be
 \frac{\partial H_z}{\partial y} = - i\omega\epsilon_0 E_x , \qquad
 \frac{\partial H_z}{\partial x} = i\omega\epsilon_0 E_y ,
 \label{eq31}
\ee
From Eqs.~(\ref{eq27}) and (\ref{eq28}) we can readily calculate the
E-polarized component of the field. Similarly, from
Eqs.~(\ref{eq30}) and (\ref{eq31}) we will obtain the H-polarized
component of the field behind the grating. Once we have these
expressions, their substitution into Eqs.~(\ref{eq22}) and
(\ref{eq23}) leads us to the corresponding EME flow lines.

\begin{figure}[t]
 \begin{center}
 \includegraphics[width=12cm]{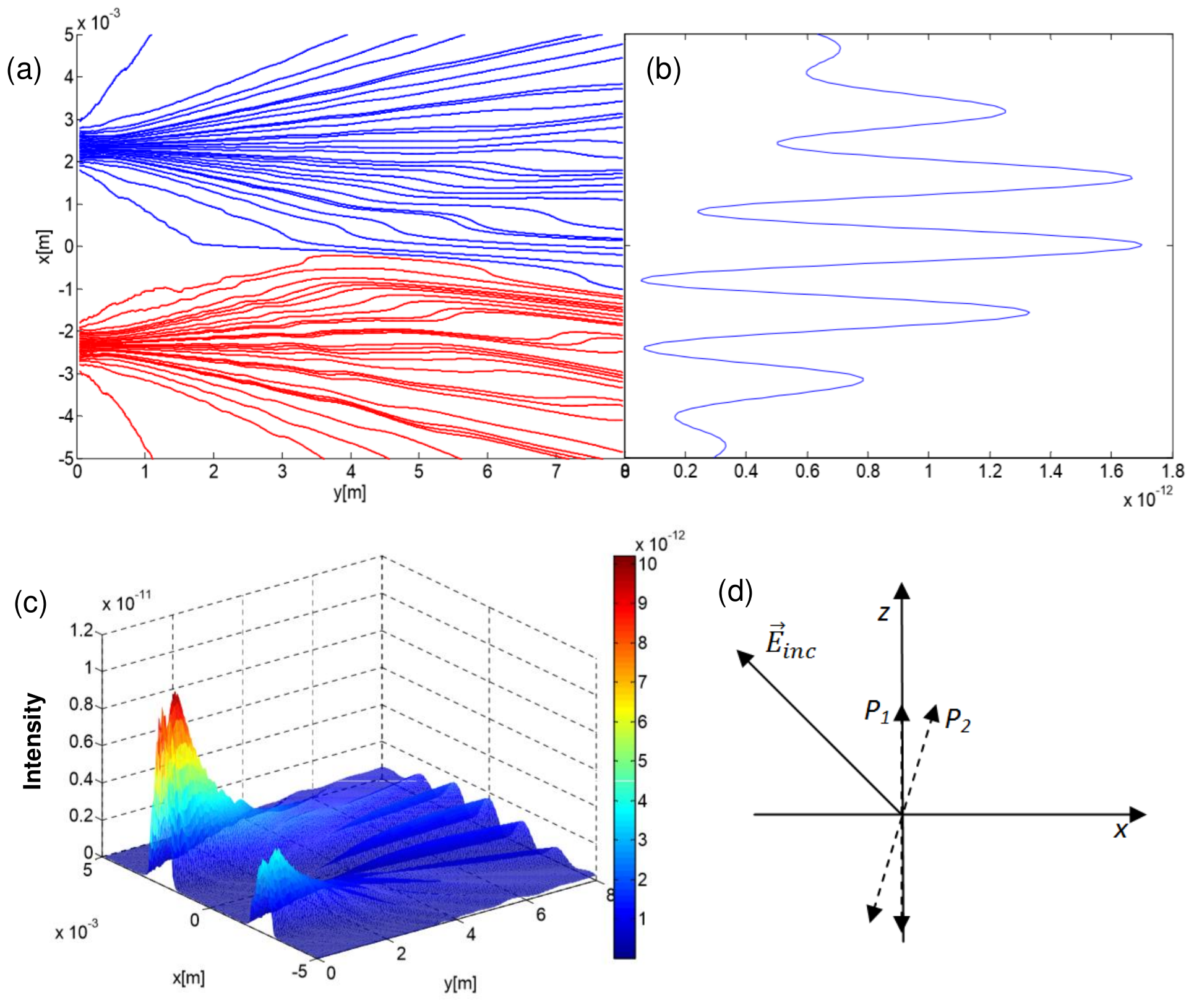}
 \caption{EME flow lines (a) and intensity distribution (b,c) behind two Gaussian
  slits followed by two polarizers and illuminated by monochromatic
  light with a wavelength $\lambda = 943$~nm. The parameters
  considered in our simulation have been taken from the experiment
  data \cite{ref13}: $\sigma_1 = \sigma_2 = 0.3$~mm,
  $\mu_1 = - \mu_2 = 2.35$~mm, and $a_1 = a_2 = 1.8\sigma_1$, and
  $\lambda = 943$~nm.
  The initial polarization is assumed to be linear, with $A=B$,
  $\varphi=0$, $\theta_1 = 0$, and $\theta_2 = \pi/10$ (d).}
 \label{fig2}
 \end{center}
\end{figure}

\begin{figure}[t]
 \begin{center}
 \includegraphics[width=11cm]{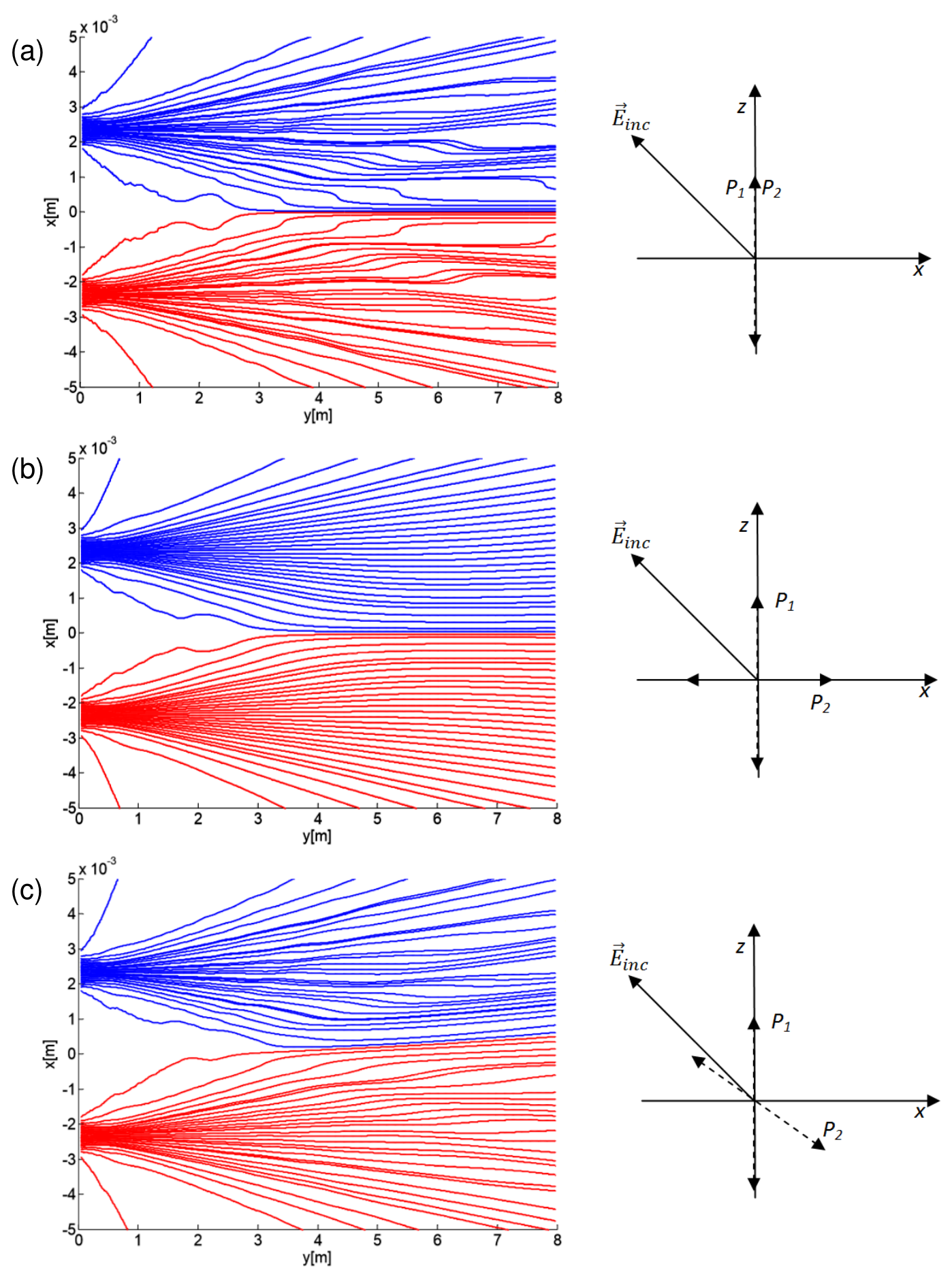}
 \caption{
  EME flow lines behind two Gaussian slits with the same parameters as
  in Fig.~\ref{fig2} and polarization angle $\theta_1 = 0$ and $\theta_2 = 0$
  (a), $\theta_1 = 0$ and $\theta_2 = \pi/2$ (b), and $\theta_1 = 0$ and
  $\theta_2 = 7\pi/10$ (c).}
 \label{fig3}
 \end{center}
\end{figure}

In Figs.~\ref{fig2} and \ref{fig3} we can observe a series of sets
of EME flow lines for different values of the polarization angles
$\theta_1$ and $\theta_2$ (the numerical details of the simulations
are provided in the corresponding captions), where the wave function
just behind each slit, and before they are acted by the polarizers,
is given by a Gaussian function:
\be
 \psi(x',0^+) = \psi_1(x',0^+) + \psi_2(x',0^+) ,
 \label{eq32}
\ee
where
\be
 \psi_i(x',0^+) = (2\pi\sigma_i^2)^{-1/4}
   e^{-(x' - \mu_i)^2/4\sigma_i^2} w(x'-\mu_i, a_i) ,
 \label{eq33}
\ee
%
with $i=1,2$ and $w(x,a)$ being the window function defined as
\be
 w(x,a) = \left\{ \begin{array}{ccc}
  1 , & \quad & x \in [-a,a] \\
  0 , & \quad & {\rm everywhere\ else}
  \end{array} \right.
 \label{eq35}
\ee

In Fig.~\ref{fig2}(a) we notice that the EME flow lines gives rise
to a more complete description of interference phenomena compared to
the description only based on intensity curves (see
Fig.~\ref{fig2}(b)). Specifically, the flow lines provide us with an
idea on how the electromagnetic field propagates through space,
while the intensity curves (see Fig.~\ref{fig2}(b)) only tell us how
much intensity is present at each point. In this sense, notice how
in the far field the maxima and minima displayed by the light
intensity correspond, respectively, to maximum and minimum values of
the density of trajectories. Actually, a slight variation in the
relative orientation of the axes of the polarizers behind the slits
gives rise to non-vanishing minima, as seen in right panel of
Fig.~\ref{fig2}(a). Only if the axes of the polarizers are oriented
along the same direction, the distribution of EME flow lines is
symmetrical with respect to the symmetry axis (see
Fig.~\ref{fig3}(a)). However, as their relative orientation
increases, as seen in the case displayed in Fig.~\ref{fig2} or in
Figs.~\ref{fig3}(b) and (c), the flow-line distribution is not
symmetrical and total fringe visibility (vanishing minima)
disappears. In this regard, the most remarkable case takes place for
mutually orthogonal polarizers (see Fig.~\ref{fig3}(b)), when the
typical oscillations of intensity disappear. It is in this case when
we talk about lack of interference. As it can be seen in any of the
figures, the distribution of EME flow lines is in compliance with
this form of intensity -it is uniform, not showing variations (see
Fig.~\ref{fig2}(b)). Apart from the fringe visibility decrease, it
can also be seen a certain phase shift, downwards for conditions
before orthogonality (see Fig.~\ref{fig2}(a)) and upwards after it
(see Fig.~\ref{fig3}(c)).

In order to check the theoretically drawn trajectories presented at
Figs.~\ref{fig2} and \ref{fig3}, we propose to experimentally
determine the average photon paths by adding differently oriented
polarizers behind the slits to the experimental setup reported in
\cite{ref13}. This is a generalization of the previous proposal by
Davidovi\'c et al. \cite{ref43} to add orthogonal polarizers in the
setup.

We consider that the proposed experiment could contribute to settle
the controversy about the influence of a potential observer on the
form of the interference pattern of two beams with different
polarizations. The standard interpretation given to the
disappearance of interference after inserting mutually orthogonal
polarizers after the slits is usually based on the Copenhagen notion
of the external observer's knowledge (information) about the photon
paths, i.e., the slit traversed by the photon on its way to the
detection screen. Proponents of the principle of complementarity
affirm that information on the path destroys the interference

Using EME-flow lines, determined from the EM field and the Poynting
vector, one arrives at another interpretation. One observes that
EME-flow lines starting from slit~1 end up on the left-hand side of
the screen, while those starting from slit~2 end up on the
right-hand side, both with presence of interference and with no
interference fringes. However, the distribution of EME flow lines is
different in each case. For the same orientation of the polarizers,
the distribution shows interference fringes (see
Fig.~\ref{fig3}(a)), while for orthogonal orientations fringes are
absent (see Fig.~\ref{fig3}(b)), in full agreement with the
Arago-Fresnel laws \cite{ref42}. Hence, the observer's information
on the slit crossed by individual photons (understood as energy
parcels) seems to be totally irrelevant regarding existence of
interference. What really matters is the form of the EME field,
which eventually influences both the distribution and the topology
of the trajectories. For values of the mutual angle between
polarizers with the interval  the symmetry of the interferometer
setup is broken. Consequently, the intensity distribution is not
symmetrical (see Figs.~\ref{fig2}(a) and (b)). Accordingly, the
distribution of trajectories is not symmetrical either, as seen in
Figs.~\ref{fig2}(a) and \ref{fig3}(c). Furthermore, a certain number
of trajectories starting at slit~1 (covered by a polarizer that
transmits more energy than the polarizer allocated behind slit~2)
cross the symmetry axis, ending up on the right-hand side (see
Fig.~\ref{fig2}(a)). Conversely, a certain number of trajectories
starting at slit~2 may cross the symmetry axis, ending up on the
left-hand side, if the polarizer allocated behind this slit
transmits more energy than the polarizer behind slit~1, as seen in
Fig.~\ref{fig3}(c).


\section{Quantum Particle Trajectories - Flow Lines of
Quantum-Mechanical Currents}
\label{sec4}

As it has been pointed out, sound waves are described in the
framework of mechanics of continuum, even though the medium through
which they propagate is composed of discrete objects, such as atoms
or molecules, for instance. In this sense, the pressure, as a
fundamental physical quantity in the description of sound waves, is
a continuous function of space and time. This description is well
accepted and experimentally verified. Nobody opposes or criticizes
such a combination of discreteness and continuity.

In quantum mechanics, wave function $\Psi({\bf r},t)$ synthesizes
continuity (wave properties) and discreteness (particle properties)
in the quantum realm. But, differently from pressure in classical
hydrodynamics, the physical meaning of the wave function has been
open to debate since the very inception of quantum mechanics (and,
by extension, also quantum optics).What is the physical meaning of a
wave function associated with an individual electron, neutron, atom,
molecule or photon? This is a longstanding question that has been
and is still looking for an answer.

It is within this context where we consider very useful taking into
account the direct analogy between the quantum mechanical current
density,
\be
 {\bf J}({\bf r},t) = \frac{\hbar}{2im}
  \left[ \Psi({\bf r},t) \nabla \Psi^*({\bf r},t)
      - \Psi^*({\bf r},t) \nabla \Psi({\bf r},t) \right]
  = | \Psi({\bf r},t) |^2 {\bf v}({\bf r},t) ,
 \label{eq36}
\ee
and the Umov vector for sound waves, described by Eq.~(\ref{eq11}),
on the on hand, and Poynting vector for electromagnetic waves,
defined by Eq.~(\ref{eq23}), on the other hand.

A more complete or, at least, more robust picture of sound wave
interference is obtained by considering flow lines determined by the
Umov vector \cite{ref28,ref29}. The same happens in the case of
photon interference when considering electromagnetic energy flow
lines \cite{ref12,ref38} determined by the Poynting vector
(\ref{eq23}). Analogously, a more complete picture of quantum
interference with massive particles arises after considering flow
lines of quantum mechanical probability density
\cite{ref24a,ref24b,ref26,ref33,ref36,ref38,ref43,ref46,ref47,ref48}. These
more complete pictures might contribute to the resolutions of
dilemmas and paradoxes related to the question of the physical
meaning of the photon wave function as well as for the quantum
mechanical wave function of nonzero mass particles.

Using quantum mechanical current density one obtains an objective
interpretation of interference phenomena as a process of
accumulation of single particle events, as confirmed in experiments
with beams of one per one, electron \cite{ref49}, neutron
\cite{ref50} atom and molecule \cite{ref51}. Intensity curves
evaluated by taking modulus square of a wave function describe only
the final distribution, obtained after accumulation of many
particles. Intensity curves do not explain the distributions of
tracks of a small number of quantum objects.


\section{Concluding Remarks}
\label{sec5}

The use of flow lines allows us to get a more complete understanding
of wave propagation and interference phenomena in classical as well
in quantum physics. This follows from the simulations and analyses
of the flow lines associated with sound waves and electromagnetic
field in a typical interference device, namely a double slit
grating. In this way, larger or smaller values of flow-line
densities are directly related to the common notions associated with
the superposition principle of constructive and destructive
interference, respectively.

In our opinion, the combined analysis of the propagation and the
evolution of flow lines for sound waves will be helpful in the long
standing debate about the interpretation of the quantum mechanical
wave function in spite of the different nature of these two kinds of
waves. Due to the analogy between classical and quantum interference
phenomena, the ability to perform in a simpler fashion experiments
with classical waves that mimic quantum behaviors, should render
important insight on quantum systems. Notice that sound waves are
described by variables that are continuous functions of space and
time, although gases and fluids, which constitute the material
substrate along which the wave propagates, are composed of
individual, discrete objects, e.g., atoms and molecules. In the same
way, trajectories in the quantum wave function do not necessarily
need to account for the individual motion of a (quantum) particle,
but provide us information on how probabilities flow in
configuration space, and therefore how averaged swarms of identical
particles travel throughout such a space.

Based on the results presented here and the recent experiment
performed by Kocsis et al. \cite{ref13}, we would also like to
propose measuring average photon trajectories behind a two-slit
grating covered by two polarizers for various mutual angles between
their polarization axes. We expect that this experiment should
render trajectories in compliance with the EME flow lines displayed
in Figs.~\ref{fig2} and \ref{fig3}. Experiments of this kind should
contribute to render some light on the longstanding debate on the
influence of a potential observer on the form of the interference
pattern of two beams with different polarizations.


\section*{Acknowledgments}
\pst
Support from the Ministry of Education, Science and Technological
Development of Serbia under Projects No. OI171005 (MB), OI171028
(MD) and III45016 (MB, MD), and the Ministerio de Econom{\'\i}a y
Competitividad (Spain) under Project No. FIS2011-29596-C02-01 (AS)
as well as a ``Ram\'on y Cajal'' Research Fellowship with
Ref.~RYC-2010-05768 (AS) is acknowledged.


\end{document}